\documentstyle[12pt,epsf,amssymb,citesort,epsfig]{article} 
\def\eq#1{Eq.~(\ref{eq:#1})}
\def\fig#1{Fig.~(\ref{fig:#1})}
\def\pt{p_T}
\def\sec#1{Section~\ref{sec:#1}}

\begin{document}
\def\om{\omega}
\def\Om{\Omega}
\def\mn{\mu\nu}
\def\al{\alpha}
\def\BM#1{{\boldmath
\mathchoice{\hbox{$\displaystyle#1$}}
            {\hbox{$\textstyle#1$}}
            {\hbox{$\scriptstyle#1$}}
            {\hbox{$\scriptscriptstyle#1$}}}}

\newcommand{\pr}[3]{{Phys.\ Rev.} {\bf #1} (19#2) #3}
\newcommand{\np}[3]{{Nucl.\ Phys.} {\bf #1} (19#2)~#3}
\newcommand{\pl}[3]{{Phys.\ Lett.} {\bf #1} (19#2) #3}
\newcommand{\beq}{\begin{equation}}
\newcommand{\eeq}{\end{equation}}
\newcommand{\beqa}{\begin{eqnarray}}
\newcommand{\eeqa}{\end{eqnarray}}
\newcommand{\md}{\mbox{d}}

\begin{flushright}
\hfill{\small MSUHEP-11004} \\
\hfill{\small hep-ph/0112191}
\end{flushright}
\begin{center}

{{\large\bf Differential Distributions for NLO 
Analyses of Charged Current Neutrino-Production of Charm}}

\vspace{10pt}

 S.~Kretzer$^1$, D.~Mason$^2$, F.~Olness$^3$

\vspace{10pt}

{\small\it $^1$Department of Physics and Astronomy, Michigan State
University, East Lansing, MI 48824}\\

\vspace{10pt} {\small\it $^2$ Department of Physics, University of Oregon, 
              Eugene, OR 97403}\\

\vspace{10pt} {\small\it $^3$ Department of Physics,  
Southern Methodist University, 
      Dallas, TX 75275-0175}

\end{center}

\vspace{24pt}

\begin{abstract}
 \noindent
Experimental analyses of charged current deep inelastic charm
production -- as observed through dimuon events in neutrino-iron
scattering -- measure the strangeness component of the nucleon sea.  A
complete analysis requires a Monte Carlo  simulation to account
for experimental detector acceptance effects; therefore, a fully
differential theoretical calculation is necessary to provide complete
kinematic information.  We investigate  the theoretical issues involved in
calculating these differential distributions at  {\it
Next-Leading-Order} (NLO).  Numerical results are presented for
typical fixed target kinematics. We present a corresponding {\tt
FORTRAN} code suitable for experimental NLO analysis.
\end{abstract}

\baselineskip=18pt
\parskip=8pt
\section{\large\bf Introduction} \label{sec:intro}
Recent 
sets of global parton distribution functions (PDFs) \cite{cteq5,grv98,MRST,zomer}
have reached a sufficiently high level of accuracy that
quantization and propagation of statistical errors have become  
important issues \cite{pdferrors}. 
It is, therefore, even  more unsettling  
that the strange quark PDF, $s(x,Q^2)$,
remains a mystery \cite{kosty}  without a fully consistent 
 picture emerging
from the comparative analysis  between neutrino and muon 
structure functions \cite{cteq1,gkr1,bgn}, 
opposite sign dimuon production in $\nu Fe$-DIS 
\cite{charm2,bazarko,ccfrlo,cdhsw,nomad,nutev,gkr1,gkr2,bgn},
or the recently measured parity violating structure
function $\Delta xF_3$ \cite{DxF3,kosty}. 
 Given the high precision of the non-strange PDF components, 
this situation for $s(x,Q^2)$ is unacceptable both in terms 
of our understanding of the nucleon structure, and for our ability 
to use precise flavor information to make predictions for present 
and future experiments. 

For extracting the strange quark PDF, the dimuon data
provide the most direct determination. 
The basic channel is the weak charged current process 
$\nu s \to \mu^- c X$ 
with a subsequent charm
decay $c \to \mu^+ X^{\prime}$. 
These events provide a direct probe of the  $s W$-vertex, and
hence the strange quark PDF.\footnote{In contrast,
single muon production only provides indirect information about 
$s(x,Q^2)$ which must then be extracted from 
a linear combination of structure functions
in the context of the QCD  parton model.}
For this reason, fixed-target neutrino dimuon production will provide
a unique perspective on the strange quark distribution of the nucleon
in the foreseeable future.  
Besides, HERA provides a large dynamic range in $Q^2$ for 
the CP conjugated process 
$e^- {\bar s} \rightarrow {\bar \nu} c$ 
which is valuable for testing the underlying QCD
evolution \cite{kstr,ref:barone}. 
Within the HERMES experimental program \cite{hermes}
the flavor structure of the polarized
and unpolarized sea are studied from semi-inclusive DIS
where DIS Kaon production has obvious potential
to probe strangeness. Thus, HERA and HERMES can complement  
fixed-target neutrino dimuon data with 
information at different energies
and from different processes;
therefore, neutrino DIS serves for now as an important
benchmark process to perform rigorous 
and refined comparisons between the
experimental data and the theoretical calculations.
In the long run, a 
high luminosity neutrino factory
could, of course, considerably
raise the accuracy of present day 
information from $\nu$ DIS \cite{nufac}. 

The theoretical calculations of inclusive charged current 
charm production have been carefully studied in the literature 
\cite{gottschalk,gkr1,acot,TR,sacot,kretzer,bn,kstr}.
Additionally, the charm fragmentation spectrum has also been calculated
in detail \cite{gkr2,kstr}. 
 While inclusive calculations are sufficient for many tasks, 
a comprehensive analysis of the experimental data at NLO 
requires additional information from the theoretical side. 
 In charged current  $\nu$-Fe  charm production, 
the  detector acceptance depends on the full range of kinematic variables: 
$\{ x, Q^2,z,\eta \}$.\footnote{Azimuthal $\phi$-dependence is controlled 
to be flat by 
rotational symmetry around the axis of the boson-target frame. 
It is, therefore, not indicated.}
Here,  $x$ is the Bjorken-$x$, 
$Q$ is the virtuality of the $W$-boson, 
$z$ is the  scaled energy of the charm  after fragmentation,
and   $\eta$ is   charm rapidity.
 The theoretical task is, {\it mutatis mutandis}, not too different
from the extraction of the neutral current charm structure function
$F_2^c$ as performed by the HERA experiments \cite{heracharm}; this
analysis uses the theoretical calculation of the differential cross
section \cite{hvqdis} to extrapolate into regions of poor acceptance.

The organization of this paper is as follows. 
In  \sec{exp}, we discuss the key factors 
that influence the acceptance of the experimental detector. 
 In Section \ref{sec:theor}, we review the theoretical calculation of the 
fully differential cross section at NLO in QCD \cite{gkr2}. 
 In \sec{num}
we present numerical  results for typical fixed target kinematics.
 Finally, in \sec{conclusions}, we draw our conclusions.

\section{\large\bf Experimental Environment: $\nu Fe$ DIS}
\label{sec:exp}

Dimuon events from neutrino charged current charm production can provide a
clear set of events from which to study the strange sea.  Their signature in a
detector is a pair of oppositely charged muons and a hadronic shower
originating from the same vertex.  The second muon is produced in the
semileptonic decay of the charmed particle.  Detector geometry and applied cuts
affect how these events are reconstructed, and must be corrected for when
making measurements of the underlying charm production.  These corrections
require a cross section differential in variables in addition to the typical
$E_{\nu}, x$, and $Q^2$ or $y$ of the charged current cross section. 

To seperate dimuon events from backgrounds a minimum energy requirement must be
applied to the muon from charm decay.  For the muon to be visible at all, it
must first be energetic enough that it travels further in the detector than the
particles which make up the hadronic shower.  The background from muons from
nonprompt decays of pions and kaons within the shower is large at low energy,
and must also be reduced.  Typically a cut on the charm decay muon's energy of
a few GeV is applied to guarantee it be reconstructable, and to reduce the
nonprompt backgrounds.  The energy of the decay muon depends on the energy of
the charmed meson from which it decays, and therefore depends on the
fragmentation parameter $z$. In NLO, it also depends on the rapidity of the
charmed quark, which maps onto the $W$-parton CMS scattering angle.  A charmed
meson with low $z$, and/or low rapidity will be less likely to pass a cut on the
decay muon's energy than one with high $z$ and maximal\footnotemark[11]
rapidity.  In addition,
the linear construction of typical fixed target neutrino detectors can
adversely affect the reconstruction of decay muons with large scattering
angles.  Considering these effects, it is important that the charm production
cross section's dependence on both $z$ and charm rapidity be understood to be
able to model dimuon events and their detector acceptance properly.

\begin{figure}[t]
\vspace{-1cm}
\epsfig{figure=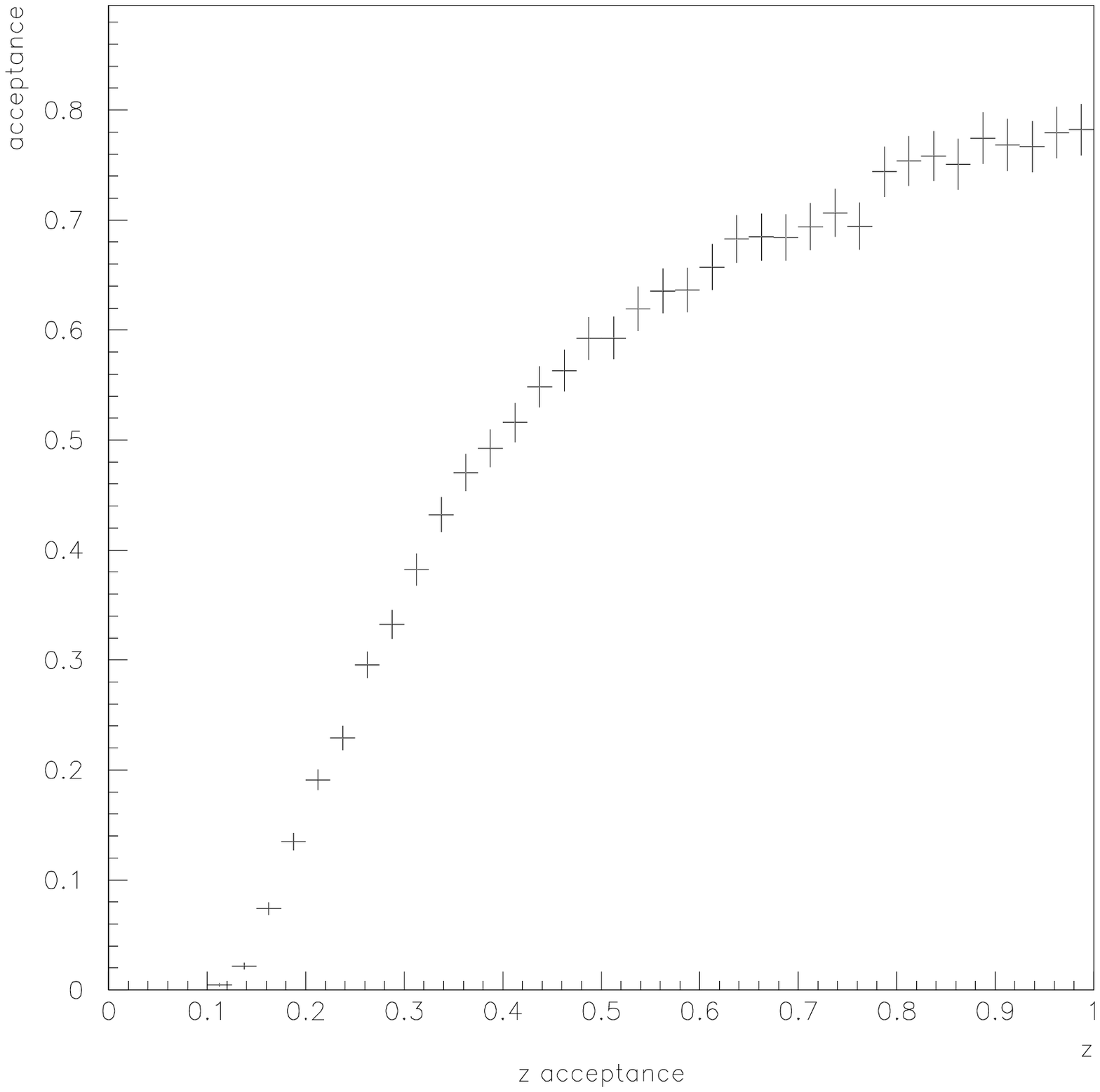,width=8cm}
\epsfig{figure=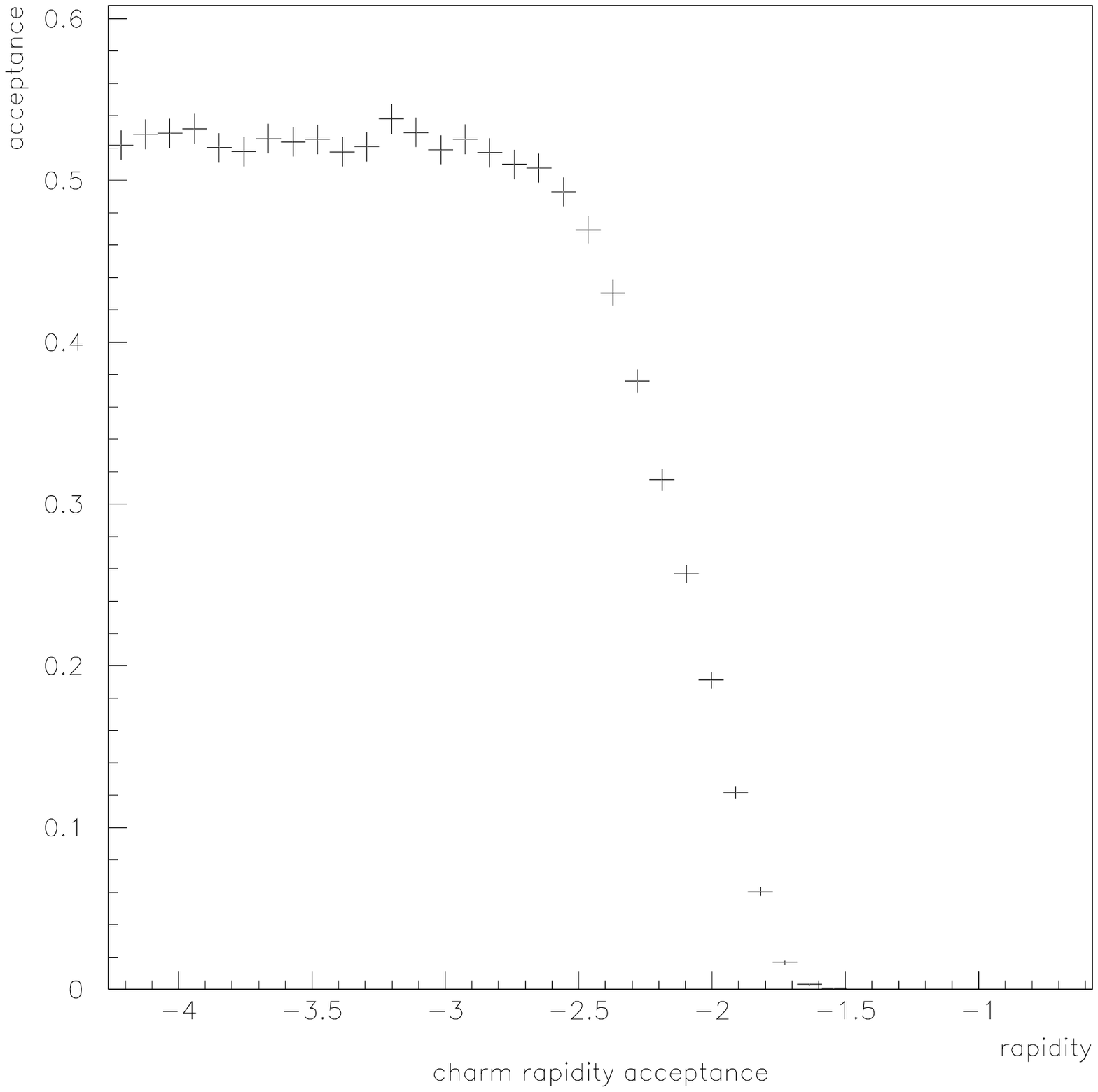,width=8cm}
\caption{Relative acceptance ${\cal{A}}(z,\eta )$ 
induced from typical kinematic cuts 
(as detailed in the text)
on the decay-muon for
$E_\nu = 80\ {\rm GeV}$, $x=0.1$, $Q^2 = 10\ {\rm GeV}^2$.
\label{fig:acc}}
\end{figure}

Fig.~\ref{fig:acc} illustrates the effect of a decay muon energy and 
angle cut on $z$ and
rapidity acceptance at a particular $E_{\nu}, x$, and $Q^2$ point.  A toy monte carlo
was constructed that generated charmed mesons with flat distributed $z$ and charm
rapidity, and decayed them into muons following \cite{bp}.  Detector smearing
effects were not included.  The figure shows muon acceptance after a 5 GeV cut
on energy, and a maximum scattering angle cut of 0.250 radians, as a function
of $z$ and charm rapidity for $E_{\nu}=80\ {\rm GeV}$, $x=0.1$, 
and $Q^2=10\ {\rm GeV}^2$.  
The acceptance is a
smooth function in both kinematic variables, but flattens out as $z$ approaches 1
and rapidity approaches its maximum value.  To properly correct for acceptance
the cross section must be differential where the slope is nonzero, but in regions
where the acceptance is flat, the dependence on these variables may be safely
integrated out; i.e. 
we conclude that the 
integrated acceptance correction
\begin{equation}
\int d z d \eta 
\ \left[
1 - {\cal{A}}(z,\eta )
\right] \frac{d \sigma}{dx\ dQ^2\ dz\ d \eta}
\end{equation}
does not resolve any local deviations of $d \sigma$
around some point ${z_0, \eta_0}$ within a range
where $[\partial {\cal{A}}(z,\eta ) / d z ] \times (z - z_0)$
and $[\partial {\cal{A}}(z,\eta ) / d \eta ] \times (\eta - \eta_0)$
are smaller than the typical MC accuracy. In our example,
these conditions are safely met within a range of, say,
1 unit in rapidity and 0.2 units in $z$ away from the phase space 
boundary.  

\section{\large\bf Differential Distributions at NLO}
\label{sec:theor}

Recorded charged-current charm production rates must be corrected for the
detector acceptance which, as discussed above, depends on the full range of
kinematic variables:
$\{ x, Q^2, z, \eta \}$.\footnotemark[2]
 Therefore, we must obtain the NLO theoretical cross section which is
completely differential in all these variables.

 While $x$ and $Q^2$ are fixed by the energy and angle of the scattered
muon, the variables $z$ and $\eta $ relate to the
center-of-mass-system (CMS) phase space of the hard partonic
scattering event, which at {\it Leading-Order} (LO) is $W^+ s^{\prime}
\rightarrow c$.
 At NLO, the quark-initiated process ($W^+ s^{\prime} \rightarrow c$)
receives virtual and real corrections, and  we
encounter a new gluon-initiated process, $W^+ g \rightarrow c
\bar{s}^{\prime}$.
Neglecting bottom contributions,\footnote{We 
can safely neglect the bottom quark-initiated contributions as these are 
suppressed both due to the bottom-quark mass, and also by the  
Cabibbo-Kobayashi-Maskawa (CKM) matrix element.}
 we denote the  CKM-rotated weak eigenstate with $s^{\prime}$, 
which is defined as:
\begin{equation}
s^{\prime} = \left|V_{s,c}\right|^2\ s + \left|V_{d,c}\right|^2\ d 
\end{equation}
Similarly, we will use
\begin{equation} 
g^{\prime}\equiv  g \left(
\left|V_{s,c}\right|^2 + \left|V_{d,c}\right|^2 \right)   
\end{equation}
to denote its QCD evolution partner, {\it i.e.}, 
$d s^{\prime} / d \ln Q^2
= s^{\prime} \otimes P_{qq} + g^{\prime} \otimes P_{qg}$.

The NLO 2-particle  partonic phase space in $N=4+2 \epsilon$ dimensions 
(appropriate for a single charm quark of mass $m_c$, plus a massless gluon or
strange quark) is: 
\begin{eqnarray} 
\int d{\mathrm{PS}_2} 
\label{eq:dps2nohat}  
&=& \frac{1}{8\pi}\ \frac{s-m_c^2}{s}
\ \frac{1}{\Gamma(1+\varepsilon)}
\ \left[ \frac{(s-m_c^2)^2}{4\pi s} 
\right]^\varepsilon\ \int_0^1\ \left[{\hat{y}}(1-{\hat{y}})
\right]^\varepsilon\ d{\hat{y}} 
\end{eqnarray}
Note that there is  only one  independent variable;\footnote{To 
count the degrees-of-freedom, we have
$2 \times 4$ momentum coordinates,
minus $2$ mass shell conditions, 
minus $4$ energy momentum conserving conditions,
minus $1$ rotational symmetry around the CMS axis.}
 In \eq{dps2nohat}, this  has been chosen  to be 
${\hat{y}}\equiv (1+\cos \theta^\ast )/2$, with $\theta^\ast$ being the 
$W^{\pm}$-parton CMS scattering angle.

The expression for the phase space in  \eq{dps2nohat}  leaves
us two options for computing the cross section: i) we can either
integrate over the entire phase space, or ii) we can compute a
singly-differential distribution.  

\subsection{Singly Differential Distributions}

We can construct any singly-differential cross section as follows. 
If the variable $\Xi$ stands for any 
kinematic variable
that can be expressed through $\hat{y}$ and partonic CMS energy  $\hat{s}$, 
 we can easily obtain the differential cross
section with respect to $\Xi$ with the appropriate Jacobian via 
the relation: $d\sigma/d \Xi = (d\sigma/d {\hat{y}})\ (d{\hat{y}}/d\Xi)$.
 As above,  ${\hat{y}}\equiv (1+\cos \theta^\ast )/2$ and 
the partonic CMS energy is given by: 
\begin{equation}
{\hat{s}} = \frac{Q^2}{\xi^{\prime}}\ \left(
1 - \xi^\prime + \frac{m_c^2}{Q^2} \right)\ \ \ , 
\end{equation}
with $\xi^\prime$ defined below in (\ref{eq:xip}). 
In our case, we consider a distribution
singly-differential in $\zeta$,
the partonic fragmentation variable
\begin{equation}
\label{eq:zeta}
\zeta \equiv \frac{p_c \cdot P_N}{q\cdot P_N}
= \frac{{\hat{y}} ({\hat{s}} - m_c^2)+m_c^2}{{\hat{s}}}
\end{equation} 
which reduces to the scaled energy of the unobservable (colored)
charm quark, 
$\zeta=E_c / E_W$, in the target rest frame. After fragmentation
the observable (colorless)
charm hadron (collectively labeled a $D$-meson)
carries some smaller energy $z=E_D / E_W < \zeta$.

The unobservable variable $\zeta$ is convoluted with a fragmentation
function  to compute the observable CC DIS  hadronic charm structure functions:
\begin{eqnarray}  \nonumber
{\cal{F}}_i^c(x,z,Q^2) &=& s^{\prime}(\xi,\mu^2_F)\ D_c(z) \\ \nonumber &+&
\frac{\alpha_s(\mu^2_F)}{2\pi}
\int_{\xi}^1 \frac{d\xi^{\prime}}{\xi^{\prime}} 
\int_{\max (z,\zeta_{\min})}^1 \frac{d\zeta}{\zeta} 
\left[H_i^q(\xi^{\prime},\zeta,\mu^2_F,\lambda)\ s^{\prime}(\frac{\xi}{\xi^{\prime}},\mu^2_F)
\right. \\ \label{eq:diff}
&+& \left.  
H_i^g(\xi^{\prime},\zeta,\mu^2_F,\lambda)\ g^{\prime}(\frac{\xi}{\xi^{\prime}},\mu^2_F)
\right] D_c(\frac{z}{\zeta}) \ \ ,
\end{eqnarray}
 In \eq{diff}, 
$D_c(z)$ is the  fragmentation function 
(often parameterized in the form of Peterson {\it et al.} \cite{peterson}), 
$z$ is the  scaled charm hadron energy,
$\xi=x(1+m_c^2/Q^2)$,
$\zeta_{\min} = m_c^2 / {\hat{s}} 
= (1-\lambda ) \xi^{\prime} / (1-\lambda \xi^{\prime})$
and $\lambda = Q^2 / (Q^2+m_c^2)$.

The outer convolution integral  over the variable
\begin{equation}
\label{eq:xip}
\xi^{\prime} = \frac{Q^2}{2p_{s,g}\cdot q}\ 
\left(1+\frac{m_c^2}{Q^2}\right)=\frac{Q^2+m_c^2}{{\hat s}+Q^2}  
\end{equation}
folds in the quark and gluon PDFs.
We take the factorization scale $\mu_F$ used in the PDFs
equal to the renormalization scale  $\mu_R$  used in $\alpha_s(\mu_R)$. 
 We find it convenient to  normalize  our  ${\cal{F}}_i^c$ 
relative to the conventional 
CC DIS structure functions $F_i^c$ as follows:${\cal{F}}_1^c \equiv F_1^c$, 
${\cal{F}}_3^c \equiv F_3^c/2$, 
${\cal{F}}_2^c \equiv F_2^c/2\xi$.
We adopt the standard normalization
of semi-inclusive structure functions
as implicitly defined through
\begin{equation} 
F_i(x,Q^2) = 
\int 
\left(\prod\limits_{\alpha \epsilon \{\Xi\}} d \alpha\right)
\ F_i\left(x,Q^2,\{\Xi\}\right) 
\end{equation}
with $\{\Xi\}$ any set of kinematic variables.

The form of the hard-scattering coefficients, $H_i^{q,g}$, can be
found in \cite{gkr2,kstr}.  These hard-scattering coefficients,
$H_i^{q,g}$, are mathematical distributions containing both
Dirac-$\delta$ and (generalized)
{\it plus} distributions arising from the soft
singularities and mass singularities present in the NLO QCD
calculation.\footnote{In this article, the term {\it distributions}
may denote either kinematic or mathematical ones; the
individual meaning will be clear from the context.}
 To map these distributions into C-numbers, smooth test functions are
required; normally, it is the parton distribution functions and the
fragmentation functions that serve as these smooth test functions.
 Convolutions over {\it both} $\xi^{\prime}$ and $\zeta$ are necessary
to guarantee C-number results for the hadronic structure function on
the left-hand-side (LHS) of \eq{diff}; the convolution over
$\xi^{\prime}$ folds in the PDFs, and the convolution over $\zeta$
folds in the fragmentation function.

In light of the above discussion, we clearly see that if we require
structure functions which are fully differential in all four
variables, such as ${\cal{F}}_i^c(x,Q^2,z,\eta)$  which is 
required to compute the experimental detector acceptance,
we will be left with unphysical Dirac-$\delta$ and {\it
plus} distributions. Let us note that
we are facing a new problem that is intimately
related to the single-charm kinematics of charged-current DIS.
There is no direct lesson to be learnt
from the pair-production kinematics
of the well-understood
neutral-current case \cite{hvqdis} where the phase space has
more degrees of freedom.   
We now turn to address this question.

\subsection{Fully Differential Distributions}

Suppose we wish to compute the structure function  
which is differential in four variables. 
As a specific example, we will choose  
${\cal{F}}_i^c(x,Q^2,z, p_T^{c,\ast} )$ 
where  $p_T^{c,\ast}$ is the transverse momentum of the 
charm particle in the boson-parton CMS frame. 
 From \eq{dps2nohat} and \eq{diff} we will find 
${\cal{F}}_i^c(x,Q^2,z, p_T^{c,\ast} )$  
contains mathematical distributions  of the form:
$\delta ( p_T^{c,\ast} - {\overline{p_{T}^{c,\ast}}} )$
and 
$1/( p_T^{c,\ast} - {\overline{p_{T}^{c,\ast}}} )_+$, 
where 
${\overline{p_{T}^{c,\ast}}} = \sqrt{(1-\zeta )(\zeta {\hat{s}} - m^2)}$.
 The $\delta$-function distributions arise from the diagrams
that have LO ($2\to1$) kinematics, whereas the plus-distributions
arise from diagrams with NLO ($2\to2$).

In principle, C-numbers are ultimately obtained  from the 
${\cal{F}}_i^c(x,Q^2,z, p_T^{c,\ast} )$ distribution 
after convolution with a smooth detector-resolution function 
in a MC program. 
 While such an approach may work in principle, 
there are many inherent difficulties 
trying to interface 
mathematical $\delta$-function distributions and plus-distributions 
with a complex detector simulation MC program. 

In the next sections, we first look at the source of the 
singularities which appear in the theoretical calculation, and then
find a means to extract the relevant distributions in an expeditious manner.

\subsection{\large\bf Resumming Soft Gluons and the Sudakov Form Factor}

As mentioned above, if we compute the fully differential 
${\cal{F}}_i^c(x,Q^2,z, \eta )$, or equivalently 
${\cal{F}}_i^c(x,Q^2,z, p_T^{c,\ast} )$,
 the final
state is over-constrained and this yields an unphysical 
delta-function distribution
in the observable cross section. While such a distribution will yield sensible
results for any integrated observable, the full differential
distribution $d\sigma / dx dQ^2 dz d\pt$ will contain
unphysical singularities.

This affliction is the limitation of our fixed-order perturbation theory;
were we able to perform an all-orders calculation, such singular 
distributions would never occur---as is appropriate for a 
physical process. In our particular case, the singular behavior of
our fixed-order perturbation theory would be resolved were we to
include   soft gluon emissions as in the familiar Sudakov
process.
 We will briefly review the Sudakov resummation of soft gluon emissions
off massless particles,
and then assess how best to work with 
${\cal{F}}_i^c(x,Q^2,z, \eta )$ for non-zero charm mass.

One of the early resummations of the QCD 
logarithms of the form $\ln Q^2/q_T^2$ arising from 
soft gluon emissions was 
performed by Dokshitzer, Diakonov and Troian (DDT) \cite{Dokshitzer:1980hw}. 
They obtained the leading-log Sudakov form factor which can be 
written generically as:
$$
S(q_T,Q) = exp\left\{
-\int_{q_T^2}^{Q^2} 
\frac{d\mu^2}{\mu^2} \ 
\frac{\alpha_s(\mu)}{\pi} \
\left(
A  \ln\frac{Q^2}{\mu^2} + B
\right)
\right\}
$$
where $q_T=-p_T / z$ is the transverse momentum of a massless particle
produced in DIS divided by fragmentation-$z$. 
The net result of this  Sudakov form factor is to smear the $\pt$ 
distribution from a singular delta-function, 
to a more physical form--often taken to be a Gaussian.\footnote{There 
are a number of different implementations of this formalism.  In
the Collins-Soper-Sterman (CSS) formalism \cite{Collins:1985kg}, 
this process is separated
into a perturbative and a non-perturbative Sudakov form factor.  The
perturbative Sudakov form factor (${\cal S}^{P}$) represents a formal
resummation of the soft-gluon emissions.  The non-perturbative
Sudakov form factor (${\cal S}^{NP}$) is introduced to deal with the
singularities in the perturbative coupling at $\alpha_s(\mu)$ at
$\mu=\Lambda_{QCD}$ \cite{Meng:1996yn,yuan}.  Separately, 
Contopanagos and Sterman \cite{Contopanagos:1994yq}
developed a
Principal Value Resummation which provides a prescription for
regularizing this singularity.  More recently, 
Ellis and Veseli \cite{Ellis:1998ii}
derived a technique that operates purely in $q_T$ without resorting to
a Bessel transform to b-space.}

The phenomenological 
approach we will take in this paper is divided into two steps. 

\begin{enumerate}

\item
 We regularize the NLO calculation of
${\cal{F}}_i^c(x,Q^2,z, \eta )$
 to provide a numerical distribution free of
delta-functions and plus-distributions.  

\item
 This result is input to a Monte Carlo (MC) calculation where 
the additional effects, including iterated soft gluon emmisions,
is modeled by a Gaussian distribution
that has been fit to data.

\end{enumerate}
To ensure that we have an accurate representation of the differential
distribution suitable for evaluating the experimental acceptance and
extracting the strange-quark PDF, we need to verify that the net
smearing due to step 1) and step 2) combine to yield the smearing
measured by experiment.

The transverse momentum of charmed particles about the direction of
the charm quark has been measured by the LEBC EHS collaboration
\cite{Aguilar-Benitez:1983jc}, and
by the neutrino emulsion experiment FNAL-E531 \cite{Ushida}.
They fit to a Gaussian distribution:
$$
\frac{dN}{d\pt^2} \sim e^{- \pt^2/\langle \pt^2 \rangle}
$$
with $\langle \pt^2 \rangle$ on the order of $\sim 900$ MeV. 

This observation implies that the details of the specific
regularization procedure are inconsequential; they can simply be
compensated by adjusting the smearing of the Monte Carlo so that the
net smearing (regularization plus Monte Carlo) yield the experimental
transverse momentum distribution.  Effectively, this parameterizes the
gluon resummation of the Sudakov form factor.
In practice, we can proceed along items {\bf a)} or {\bf b)}
of Section \ref{sec:fully} and
adjust the regularization parameters by either 
varying the $\langle \pt^2 \rangle$ in the $G$ function in \eq{gaussian}
below, 
or by varying the bin width differential distribution. 
This cross-check ensures that we have an accurate representation of the 
differential distribution suitable for evaluating the experimental acceptance
and extracting the strange-quark PDF.

We now discuss the implementation of this procedure in the 
following section.

\subsection{\large\bf Implementing the Fully Differential Distribution }
\label{sec:fully}

As demonstrated above, the singular distributions are 
nothing but an unphysical artifact of
regularized perturbation theory; for any physically observable quantity,
these singularities will be smeared  by soft gluon emission 
to yield physical C-number distributions \cite{pavel,matteo}.
 As the theoretical machinery of soft gluon resummation 
is not fully developed for semi-inclusive DIS heavy quark production, 
we will use a phenomenological approach.\footnote{Note, the 
application of the Sudakov resummation to the case of 
semi-inclusive DIS heavy quark production is under current 
investigation; such an approach would be of use in a future
very-high-statistics experiment where the detailed $\pt$-distribution 
could be measured.}
 There are two viable options that we have investigated. 

\begin{itemize}
\item[a)] 
{\bf Gaussian $\pt$-smearing:}
We may smear the singular peak which appear 
in the differential structure functions.

\item[b)] \label{b}
{\bf Binning:}
We may map the singular distributions onto C-numbers distributions 
by integrating over bins
which reflect the finite resolution of the experimental detector.

\end{itemize}
We examine these possibilities in order. 

\subsubsection{Gaussian $\pt$-smearing}

We can implement Gaussian $\pt$-smearing  by replacing the delta function 
$\delta(p_T^{c,\ast}-{\overline{p_{T}^{c,\ast}}})$
with a narrow Gaussian distribution 
$G(p_T^{c,\ast},{\overline{p_{T}^{c,\ast}}}, \delta p_T^{c,\ast})$
centered about the perturbative value ${\overline{p_{T}^{c,\ast}}}$,
and with a width of $\delta p_T^{c,\ast}$:\footnote{Note, 
the actual normalization of the Gaussian is more subtle than shown 
here as the integration
in $p_T^{c,\ast}$ will be constrained by kinematic limitations. 
This complication is one reason that we do not implement this 
approach in practice. }
\begin{equation}\label{eq:gaussian}
G(p_T^{c,\ast},{\overline{p_{T}^{c,\ast}}}, \delta p_T^{c,\ast})
=
{1 \over \delta p_T^{c,\ast} \ \sqrt{\pi}} \ 
 e^{
{-2(p_T^{c,\ast}-{\overline{p_{T}^{c,\ast}}})^2 \over (\delta p_T^{c,\ast})^2 }
}
\end{equation}
Once we have a smooth distribution in $p_T^{c,\ast}$, we can 
trivially switch to any other variable, $\Xi$,
by applying the proper Jacobian:\footnote{Note, 
$G(\Xi,{\overline{\Xi}},\delta \Xi )$ will not necessarily 
be a Gaussian.}
\begin{equation}\label{eq:jaco}
G(\Xi,{\overline{\Xi}},\delta \Xi ) =  
G(p_T^{c,\ast},{\overline{p_T^{c,\ast}}}, \delta p_T^{c,\ast})
\times \frac{\partial {\overline{p_T^{c,\ast}}}}{\partial \Xi}
\end{equation} 
 For non-zero $\delta p_T^{c,\ast}$, we pass C-numbers to the MC.
In the limit $\delta p_T^{c,\ast} \rightarrow 0$,
we have
$G(p_T^{c,\ast},{\overline{p_T^{c,\ast}}}, 0)
=
\delta(p_T^{c,\ast}-{\overline{p_T^{c,\ast}}})$, and the original
mathematical (singular) distributions are recovered.
In practice, a Gaussian smearing reagularization of the 
$H_i^{q,g}$ in Eq.~(\ref{eq:diff}) corresponds to a
replacement 
\begin{equation}
\label{eq:gauss}
H_i^{q,g}(\xi^{\prime},\zeta,\mu^2_F,\lambda) \rightarrow
{\overline{H}}_i^{q,g}(\xi^{\prime},\zeta,\mu^2_F,\lambda)
= H_i^{q,g}(\xi^{\prime},\zeta,\mu^2_F,\lambda) \times
G(p_T^{c,\ast},{\overline{p_T^{c,\ast}}})
\end{equation}
with ${\overline{p_T^{c,\ast}}}\equiv {\overline{p_T^{c,\ast}}}
(\xi^\prime , \zeta )$ understood as a function of $\xi^\prime$ and
$\zeta$.

\subsubsection{Finite Detector Resolution and Binning Distributions}
\label{sec:binning}

An alternate approach is to  
map the singular distributions onto C-numbers distributions 
by integrating over bins
which reflect a finite resolution of the detector.
Additionally, this technique can 
avoid passing negative weights to the MC if 
the bin size is sufficiently large enough to allow the KLN theorem 
to effect its cancellations. This point 
will be further discussed in Section \ref{sec:num}.

In practice, the binning
means that in \eq{diff}  we make the replacement
\begin{equation}
\label{eq:theta}
H_i^{q,g}(\xi^{\prime},\zeta,\mu^2_F,\lambda) \rightarrow
{\tilde{H}}_i^{q,g}(\xi^{\prime},\zeta,\mu^2_F,\lambda)
= H_i^{q,g}(\xi^{\prime},\zeta,\mu^2_F,\lambda) \times
\Theta_{\Xi} (\xi^{\prime}, \zeta )
\end{equation}
where the step function
$\Theta_{\Xi} (\xi^{\prime}, \zeta ) = 1$ 
if 
$\Xi \equiv\Xi (\xi^{\prime}, \zeta )$
is inside the bin $\Delta \Xi$ over which to integrate,  and zero otherwise.
 Note, this means that $\Theta_{\Xi} (\xi^{\prime}, \zeta )$ is subject to
the distribution prescriptions of the original $H_i^{q,g}$ as given
in Refs.~\cite{gkr2,kstr}; {\it i.e.},  
$\Theta_{\Xi} (\xi^{\prime}, \zeta )$ must
be treated with care as
a multiplicative factor to the  
non-singular ``test-function'' part of the $H_i^{q,g}$.

\subsubsection{Practical Considerations}

We will  briefly comment why we favor implementing the binning 
of choice b) as opposed to the Gaussian $\pt$-smearing. 
 For the present case of the charm transverse momentum, $\Xi=p_T^{c,\ast}$, 
(or equivalently, the charm rapidity $\Xi=\eta$), Gaussian
$\pt$-smearing leads to a integrable but delicately singular behavior
near the boundary of phase space -- which corresponds to the singular
point of the distributions -- unless we smear the $p_T^{c,\ast}$
variable with a width $\delta p_T^{c,\ast}$ as large as $\sim m_c$.

In the context of resummation of soft gluon emissions into the Sudakov
form factor, the width $\delta p_T^{c,\ast}$ should be tied to a scale
characteristic of the soft processes within the hadron, possibly with
a logarithmic growth \cite{Apanasevich:1999ki}. For fixed-target neutrino DIS,
$p_T^{c,\ast} \sim m_c$ represents a comparatively large transverse
momentum.  Additionally, there is the complication of the
normalization of the Gaussian given the constraints of the phase space
boundaries.

Consequently, while the Gaussian $\pt$-smearing may have a simple
intuitive interpretation, we find the second method of using finite
width bins to be more economical in terms of CPU time and numerical
stability.  Results for this approach with $\Xi$ identified with charm
rapidity $\eta$ will be presented in the next section.

\section{\large\bf Numerical Results}\label{sec:num}

We have previously addressed  in Section \ref{sec:theor} the complication of 
mapping mathematical distributions onto
C-number functions  by the introduction of bins.
We also encounter large and negative Sudakov logarithms close to the 
phase space boundary where, at fixed-order, 
soft single-gluon emission is enhanced. 
Let us, for illustration, 
consider the convolution of a
plus-distribution $1/(1-x)_+$, 
and a sufficiently smooth 
and monotonically decreasing
test function  $f(x)>0$. Then, the integral
\begin{equation}
\label{eq:toy}
\int^1_{x_{\min}} dx\ \frac{f(x)}{(1-x)_+}
= \int^1_{x_{\min}} dx\ \frac{f(x)-f(1)}{(1-x)}
+ f(1) \ln (1-x_{\min})
\end{equation}
will be positive for $x_{\min}$ not too close to $1$. 
As in this simplified ``toy'' case,
the Sudakov logarithms near the phase space boundary
diverge in the limit of zero bin-width 
[$x_{\min} \rightarrow 1$ in (\eq{toy})]; 
as we increase our resolution via narrow binning, we begin to 
resolve the unphysical $\delta$-functions and ``plus-distributions.''
Conversely, by using broad bins, we are effectively integrating over
enough phase-space so that the  KLN theorem ensures that
we obtain positive physical results. 

\subsection{Kinematics for Rapidity Distributions}

We now compute  the normalized differential charm production cross
section:
\begin{equation}
\label{eq:normal}
d \sigma_{\{x,y,z,\eta\}}
\equiv
\frac{d \sigma}{dx\ dy\ dz\ d \eta}\ \left[
\frac{2 G_F^2 M_N E_{\nu}}{\pi \left(1+\frac{Q^2}{M_W^2}\right)^2}\right]^{-1}
\end{equation}
We can compute $d \sigma_{\{x,y,z,\eta\}}$ from the master equation 
for the  differential
structure function, \eq{diff},  with the binning procedure defined
by \eq{theta}. 
 For our variables, we choose the set $\{  x, Q, z, \eta \}$
where  $\eta$ is  the charm rapidity
evaluated in the collinear ($p_{\perp ,W}=0$)
target rest frame.

In the partonic center of mass system, the
charm rapidity is defined by:
\begin{equation}
\label{eq:etaCMS}
\eta^\ast = \frac{1}{2} \ln 
\frac {E_c^\ast + p_{L,c}^\ast}{E_c^\ast - p_{L,c}^\ast} 
\end{equation}
and the other necessary variables are: 
\begin{eqnarray}
E_c^\ast 
&=& 
\frac{ {\hat s} +m_c^2 }{2 \sqrt{{\hat s}}} 
 \\
p_{L,c}^\ast 
&=& 
- \ \frac{ {\hat s} -m_c^2 }{2 \sqrt{{\hat s}}} \ \cos \theta^\ast
 \\
\cos \theta^\ast 
&=& 
1- 2 \frac{(1-\zeta)(1-\lambda \xi^\prime)}{1-\xi^\prime}
\end{eqnarray}
 We find the charm rapidity to be a convenient variable as 
the relation between the rapidity in  the partonic center of mass system, 
$\eta^*$ and the rapidity in  collinear
target rest frame $\eta$ are related by a simple relation: 
\begin{equation}
\label{eq:boost}
\eta = \eta^\ast + \frac{1}{2} 
\ln \left[ \frac{M_N^2}{Q^2} x 
\left(\frac{\xi}{\xi^\prime} -x\right) \right] 
\end{equation}
 We will use the collinear target rest frame as our reference frame
since it is related to the laboratory frame by a simple spatial
rotation which depends on an event-by-event basis on the kinematics of
the leptonic vertex.

\subsection{Tracking Down Sudakov Logarithms}
%

\begin{figure}[t]
\vspace*{-1cm}
\hspace*{3cm}
\epsfig{figure=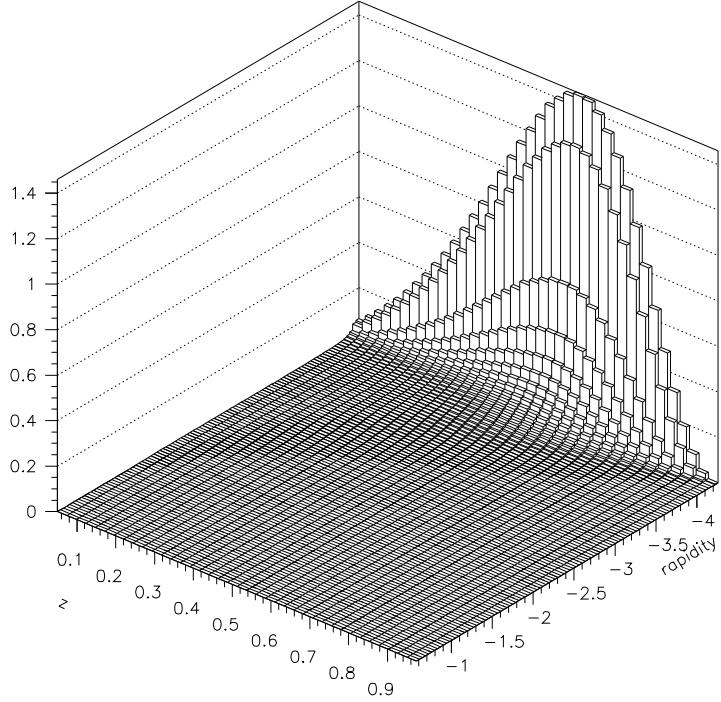,width=11cm}
\caption{Binned differential distribution for CC 
neutrino-production of charm on an isoscalar target; 
the kinematics shown are for a typical wide-band beam on a fixed
target: $E_\nu = 80\ {\rm GeV}$, $x=0.1$, $Q^2 = 10\ {\rm GeV}^2$.
\label{fig:lego}
}
\end{figure}

\begin{figure}[t]
\epsfig{figure=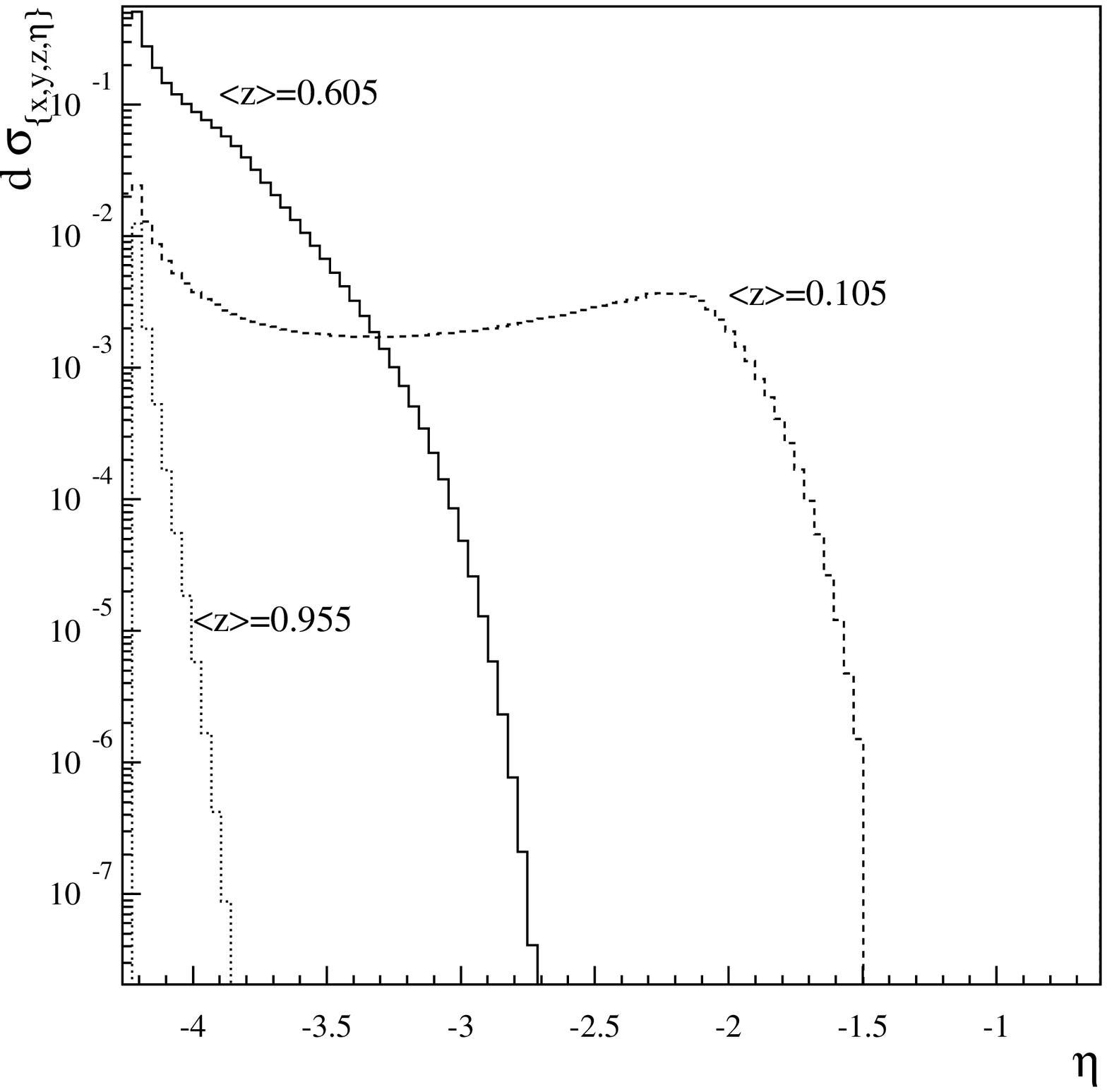,width=7.75cm}
\hspace*{0.5cm}
\epsfig{figure=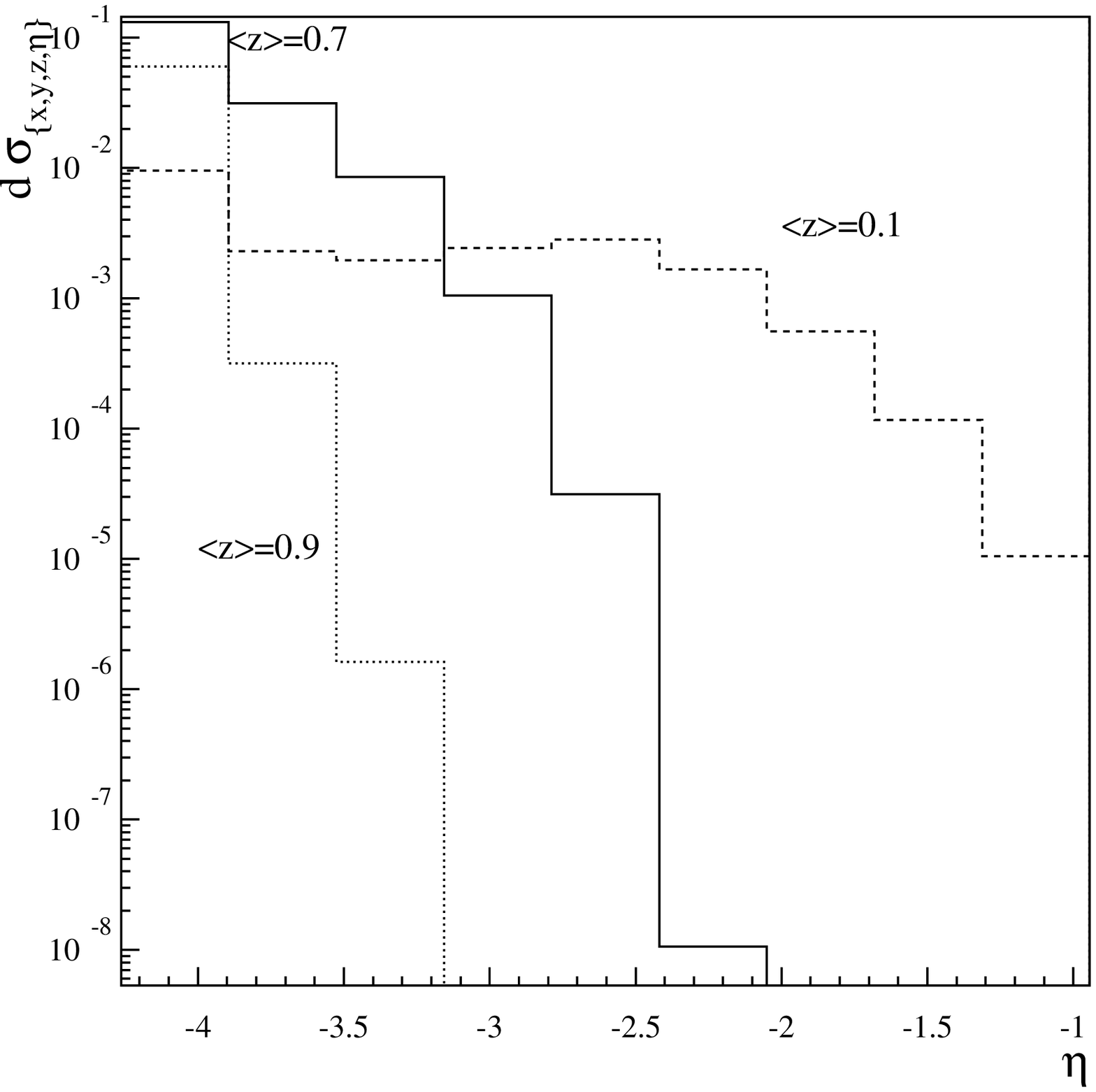,width=7.75cm} 
\caption{Binned differential distributions in 
the charm rapidity, $\eta$. Shown are
results for a fine binning (100$\times$100) in $\eta \times z$ 
(left), and a broad binning (right) of 10$\times$5. The fine-binned results
refer to $z$-bins centered around 0.105, 0.605 and 0.955, respectively, while
the broad-binned results refer to  0.1, 0.7 and 0.9.
\label{fig:2dim}
}
\end{figure}

In  \fig{lego} and \fig{2dim} we present
results for kinematics typical 
of a wide-band neutrino beam on a fixed target:  
$E_\nu = 80\ {\rm GeV}$, $x=0.1$, $Q^2 = 10\ {\rm GeV}^2$.
We plot $d \sigma$  in 2-dimensions {\it vs.} $z$ and $\eta$
in \fig{lego}, and in \fig{2dim} we show 1-dimensional slices 
for distinct bins in $z$ plotted against $\eta$.
 Comparing \fig{2dim}-a and \fig{2dim}-b, we see the effect of
the narrow- and wide-binning. 

\begin{figure}[t]
\epsfig{figure=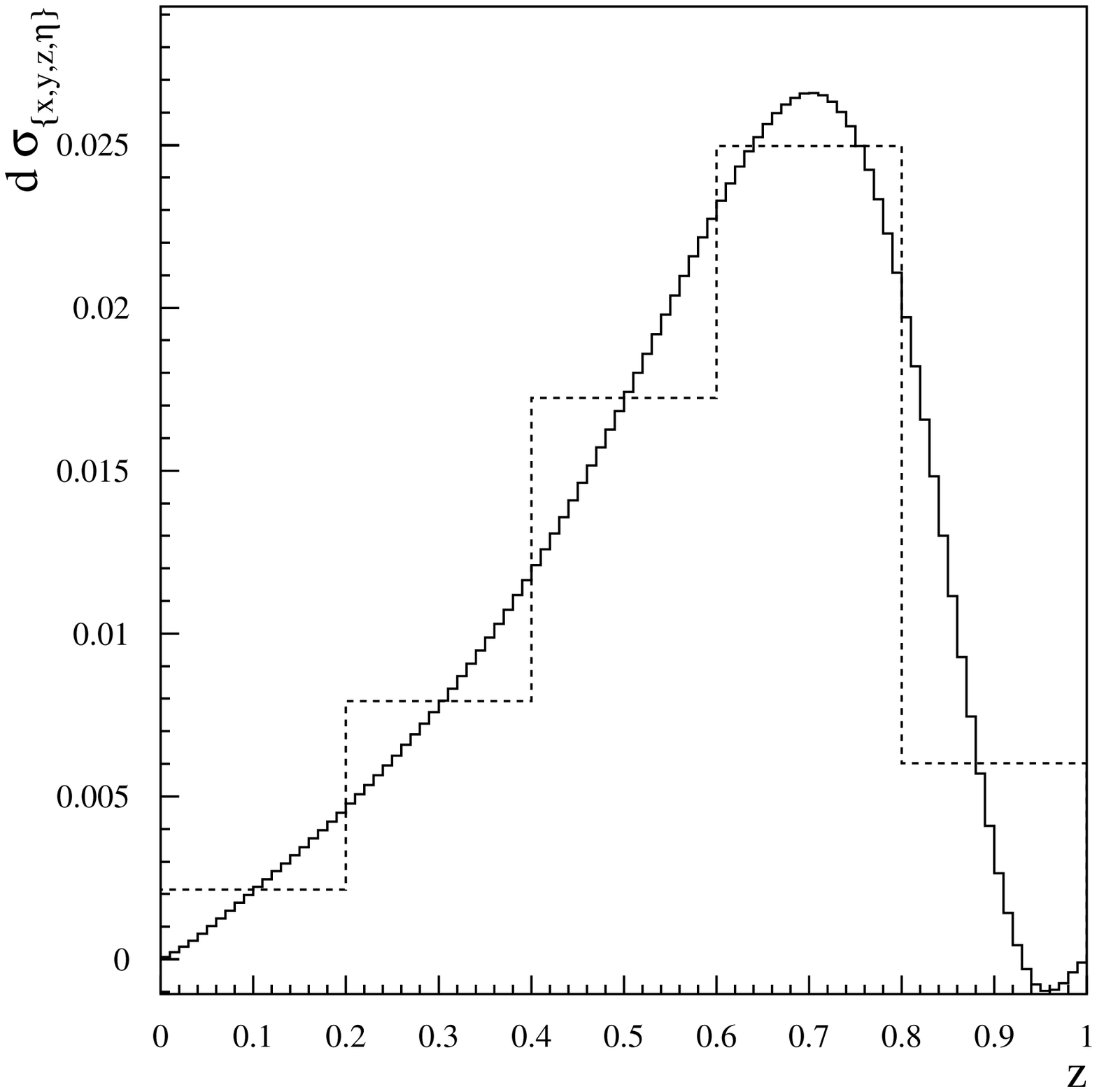,width=7.75cm}
\hspace*{0.5cm}
\epsfig{figure=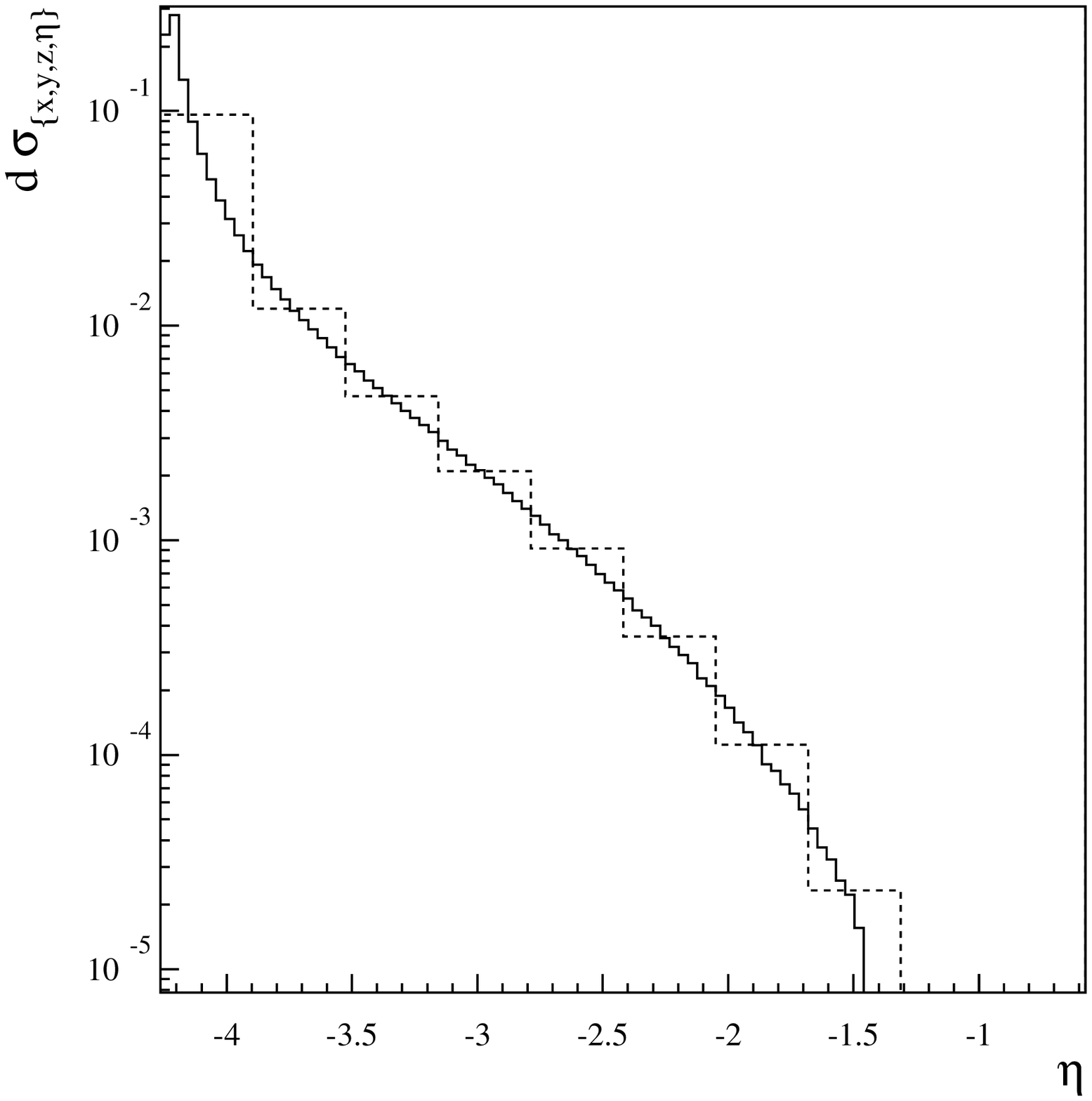,width=7.75cm} 
\caption{Binned distribution in $z$ (left: $\eta$ integrated out)
and $\eta$ (right: $z$ integrated out), each for a fine and a broad 
binning. 
\label{fig:1bin}
}
\end{figure}

 Finally, in \fig{1bin}, we show results 
for $d \sigma_{\{x,y,z,\eta ,Q^2\}}$ as defined by  \eq{normal}
for the case where either
$\eta$ or $z$ is integrated out; again, we display this for 
fine-binnings of 
1$\times$100 or 100$\times$1, and broad-binnings of
1$\times$5 or 10$\times$1.\footnote{As an important 
cross-check on these results, we verify that the binning
$\eta \times z = 1 \times n$ (rapidity integrated out) reproduces
the integrated results  in the literature  \cite{gkr2,kstr}}.

As expected, the negative Sudakov logarithms occur at large $z$ where 
we have $\eta \lesssim \eta_{\max}$ in the integrand.\footnote{
Due to our orientation of the $z$-axis along the target direction we 
denote 
$\eta_{\max} \equiv \max \left( \left| \eta \right| \right)
= \max \left( - \eta  \right)
$
} 
This behavior is clearly evident in \fig{2dim} where we see 
that in the case of fine binning (\fig{2dim}-a) and 
large $z$  (dotted curve, corresponding to $z \epsilon [0.95; 0.96]$), 
the $\eta$-distribution turns negative in the lowest
rapidity bin. 
In contrast, in the case of broad binning (\fig{2dim}-b) and 
large $z$  (dotted curve, corresponding to $z \epsilon [0.8; 1]$), 
the $\eta$-distribution remains positive throughout the full $z$-range. 
Similarly, the $z$-distribution in (\fig{1bin}-a) is rendered 
positively definite if the binning is broadened at large-$z$.
Hence, we observe that negative weights
can be easily avoided  by using sufficiently broad bins in 
a reasonable broadening of the binning

To summarize, we observe that the negative differential distributions
arising from the negative Sudakov logarithms yield unphysical singular
components present in our fixed-order result.  When sufficiently broad
bins are chosen, these singular components integrate out to yield a
positive definite binned result.  In an actual experimental analysis,
this requirement of broad bins arises naturally given the finite
detector resolution.

Our use of bins to regularize the differential distributions 
will have negligible impact on the experimental analysis 
because our bin size is small compared to the experimental detector 
resolution, even more so 
because the detector acceptance 
is a smooth function in terms of the set of kinematic variables. 
In particular, while the exact choice of bins is subject to
any experimental analysis,
the geometry of typical neutrino detectors as reflected in
the acceptance functions of Fig.~\ref{fig:acc}
demonstrates
the effective experimental bin size in $z$ and $\eta$ is 
{\underline{comfortably}} large enough for the purpose of
regularizing the  differential distributions with the binning 
technique.

To conclude this Section, 
let us note that in the lowest-$z$ curves of \fig{2dim}, one
observes -- apart from the forward peak at $\eta \lesssim \eta_{\max}$ 
typical of an $W s^\prime \rightarrow c$ event -- a 
broader backward plateau stemming from the 
scattering off a collinear (in the boson-target CMS) charm quark.
For the fixed target kinematics under consideration,
this backward peak is seen 
to be strongly
suppressed  compared 
to the forward peak. This observation
gives us confidence that one is actually measuring the
scattering off strange quarks,  with only a small dilution from
charm quarks. Hence, this verifies the  opposite
sign dimuon data provides a direct determination of 
$s(x,Q^2)$ at NLO level.

\section{\large\bf Conclusions}
\label{sec:conclusions}

We have presented  a fully differential NLO calculation of the 
neutrino-induced DIS charm production process. This calculation 
is an essential ingredient for a complete analysis of the 
dimuon data, and will allow a precise determination of the 
strange quark PDF. 
 We have demonstrated that by binning the data appropriately, 
we can interface the theoretical calculation 
(containing $\delta$-functions and ``plus-distributions'')
 directly to the experimental 
Monte Carlo analysis program. 
 We observe the enhancement of the  Sudakov logarithms
at the phase-space boundaries, and verify that these can 
be controlled with this binning method.

The fully differential distributions obtained here 
allow charged current neutrino DIS experiments
to use the complete NLO QCD result in the Monte Carlo 
data analysis.\footnote{These 
results are available as a {\tt FORTRAN} code, {\tt DISCO}.}
 These tools will allow us to extract the
strange quark PDF from the dimuon data at NLO; this information
should prove crucial to resolving the unusual behavior of the strange 
quark in the proton.

\begin{center}
\section*{\large\bf Acknowledgment}
\end{center}
The Authors would like to thank 
T.~Adams, T.~Bolton, R.~Frey, M.~Goncharov, J.~Morfin,
R.~Scalise, P.~Spentzouris, and W.-K.~Tung
for helpful discussions.
This research was supported by the
National Science Foundation under Grant PHY-0070443,
by the U.S. Department of Energy,
and by the Lightner-Sams Foundation. 
\newpage

\begin{center}
\section*{\large\bf References}
\end{center}



\begin{thebibliography}{99}

\bibitem{cteq5}    
H.~L.~Lai {\it et al.}  [CTEQ Collaboration],
Eur.\ Phys.\ J.\  {\bf C12}, 375 (2000)

\bibitem{grv98}    
M.~Gl\"{u}ck, E.~Reya and A.~Vogt,
Eur.\ Phys.\ J.\  {\bf C5}, 461 (1998).

\bibitem{MRST}    
A.D. Martin, R.G. Roberts, W.J. Stirling and R.S. Thorne,
Eur. Phys. J. C4, 463, 1998.

\bibitem{zomer}
V.~Barone, C.~Pascaud and F.~Zomer,
Eur.\ Phys.\ J.\ {\bf C12}, 243 (2000). 

\bibitem{pdferrors}
S.~Alekhin, Eur.\ Phys.\ J.\` {\bf C10}, 395 (1999);
W.~T.~Giele and S.~Keller, Phys.\ Rev.\ D {\bf 58}, 094023 (1998);
W.~T.~Giele, S.~Keller and D.~A.~Kosower, ``Parton distributions with errors,''
{\it  In *La Thuile 1999, Results and perspectives in particle physics* 255-261};
J.~Pumplin, D.~R.~Stump and W.~K.~Tung,
{\tt hep-ph/0008191}, {\tt hep-ph/0101032},  {\tt hep-ph/0101051}. 

\bibitem{kosty}
S.~Kretzer, F.I.~Olness, R.J.~Scalise, R.S.~Thorne and U.K.~Yang,
Phys.\ Rev.\ {\bf D64}, 033003 (2001).

\bibitem{cteq1}
J.~Botts {\it et al.}, CTEQ1, Phys.~Lett.~{\bf B304}, 159 (1993).

\bibitem{gkr1}
M.\ Gl\"{u}ck, S.\ Kretzer and E.\ Reya,
Phys.\ Lett.\ {\bf{B380}}, 171 (1996); {\bf{B405}}, 391 (1996) (E).

\bibitem{bgn}
V.~Barone, M.~Genovese and N.N.~Nikolaev, 
Z.\ Phys.\ {\bf C70}, 83 (1996). 

\bibitem{charm2}  
P.~Vilain {\it et al.}  [CHARM II Collaboration],
Eur.\ Phys.\ J.\  {\bf C11}, 19 (1999).

\bibitem{bazarko}  
A.~O.~Bazarko {\it et al.}  [CCFR Collaboration],
Z.\ Phys.\  {\bf C65}, 189 (1995);
A.~O.~Bazarko, Ph.D. Thesis. NEVIS-1504

\bibitem{ccfrlo} 
S.A.\ Rabinowitz {\it{et al.}} [CCFR Collaboration],
Phys.\ Rev.\ Lett.\ {\bf{70}}, 134 (1993).

\bibitem{cdhsw}
H.\ Abramowicz {\it{et al.}} [CDHSW Collaboration], C.\ Phys.\ {\bf{C15}}, 19 (1982).

\bibitem{nomad}
P.~Astier {\it{et al.}} [NOMAD Collaboration],
\pl{B486}{00}{35}.

\bibitem{nutev}
M.~Goncharov {\it et al.} [NuTeV Collaboration],
Phys.\ Rev.\ {\bf D64}, 112006 (2001). 

\bibitem{gkr2}
M.\ Gl\"{u}ck, S.\ Kretzer and E.\ Reya,
Phys.\ Lett.\ {\bf{B398}}, 381 (1997); {\bf{B405}}, 392 (1997) (E).

\bibitem{DxF3}
U.K.~Yang {\it et al.} [CCFR/NuTeV Collaboration],
Phys.\ Rev.\ Lett.\ {\bf 86}, 2742 (2001). 

\bibitem{kstr}
S.~Kretzer and M.~Stratmann,
Eur.\ Phys.\ J.\ {\bf C10}, 107 (1999). 

\bibitem{ref:barone} V.\ Barone, U.\ D'Alesio and M.\ Genovese,
in proceedings of the 1995/96 workshop on `Future Physics at HERA', Hamburg, 1996, 
G.\ Ingelman, A.\ De Roeck and R.\ Klanner (eds.), p.~102.

\bibitem{hermes}
K.\ Ackerstaff {\it{et al.}} [HERMES Collaboration],
Phys.\ Rev.\ Lett.\ {\bf{81}}, 5519 (1998);
Phys.\ Lett.\ {\bf{B464}}, 123-134 (1999). 

\bibitem{nufac}
R.D.~Ball, D.A.~Harris, K.S.~McFarland,
proceedings of NuFACT'00 ({\it International Workshop on Muon 
Storage Ring for a Neutrino Factory, Monterey, California, 22-26 May 2000}), 
{\tt hep-ph/0009223};
S.~Forte, M.L.~Mangano, G.~Ridolfi,
Nucl.\ Phys.\ {\bf B602}, 585 (2001).  

\bibitem{gottschalk}
T.\ Gottschalk, Phys.\ Rev.\ {\bf{D23}}, 56 (1981).

\bibitem{acot}  
F.~Olness, W.K.~Tung, \np{B308}{88}{813};
M.~Aivazis, F.~Olness, W.K.~Tung, \pr{D50}{94}{3085};
M.~Aivazis, J.C.~Collins, F.~Olness, W.K.~Tung, \pr{D50}{94}{3102}.

\bibitem{TR}  
R.S. Thorne, R.G. Roberts, 
\pl{B421}{98}{303};
\pr{D57}{98}{6871};
Eur.\ Phys.\ J.\ {\bf C19} 339 (2001). 

\bibitem{sacot}  
M.~Kr\"amer, F.~Olness, D.~Soper, 
Phys.\ Rev.\ {\bf D62}, 096007 (2000). 

\bibitem{kretzer}    
S. Kretzer, I. Schienbein
\pr{D56}{97}{1804}; 
\pr{D58}{98}{094035};
\pr{D59}{99}{054004}. 

\bibitem{bn}
M.~Buza, W.L.~van Neerven,
Nucl.\ Phys.\ {\bf B500}, 301 (1997). 

\bibitem{heracharm}
H1 Collab.\ and ZEUS Collab.\ (Felix Sefkow for the collaboration); 
proceedings of 30th 
International Conference on High-Energy Physics (ICHEP 2000), Osaka, Japan, 27 Jul - 2 Aug 2000; 
{\tt hep-ex/0011034}. 

\bibitem{hvqdis}
B.W.~Harris and J.~Smith, Nucl.\ Phys.\ {\bf B452}, 109 (1995). 

\bibitem{bp}
V.~Barger, R.J.N.~Phillips,
Phys.\ Rev.\ {\bf D14}, 80 (1976). 

\bibitem{peterson}
C.\ Peterson {\it et al.}, Phys.\ Rev.\ {\bf{D27}}, 105 (1983).

\bibitem{Dokshitzer:1980hw}
Y.~L.~Dokshitzer, D.~Diakonov and S.~I.~Troian,
Phys.\ Rept.\  {\bf 58}, 269 (1980).

\bibitem{Collins:1985kg}
J.~C.~Collins, D.~E.~Soper and G.~Sterman,
Nucl.\ Phys.\ B {\bf 250}, 199 (1985).

\bibitem{Contopanagos:1994yq}
H.~Contopanagos and G.~Sterman,
Nucl.\ Phys.\ B {\bf 419}, 77 (1994)
[hep-ph/9310313].

\bibitem{Ellis:1998ii}
R.~K.~Ellis and S.~Veseli,
Nucl.\ Phys.\ B {\bf 511}, 649 (1998)
[hep-ph/9706526].

\bibitem{Aguilar-Benitez:1983jc}
M.~Aguilar-Benitez {\it et al.}  [LEBC-EHS Collaboration],
Phys.\ Lett.\ B {\bf 123}, 103 (1983).

\bibitem{Ushida}
N.~Ushida {\it et al.}  [Fermilab E531 Collaboration],
Phys.\ Lett.\ B {\bf 206}, 375 (1988);
Phys.\ Lett.\ B {\bf 206}, 380 (1988); \\
N.~Ushida {\it et al.}
[Canada-Japan-Korea-USA Hybrid Emulsion Spectrometer Collaboration],
Phys.\ Lett.\ B {\bf 121}, 292 (1983).

\bibitem{pavel}
P.~Nadolsky, D.R.~Stump and C.-P.~Yuan,
Phys.\ Rev.\ {\bf D61} (2000) 014003, {\tt hep-ph/0012261}.

\bibitem{matteo}
M.~Cacciari, S.~Catani, 
Nucl.\ Phys.\ {\bf B617}, 253 (2001). 

\bibitem{Apanasevich:1999ki}
L.~Apanasevich {\it et al.},
Phys.\ Rev.\ D {\bf 59}, 074007 (1999)
[hep-ph/9808467].

\bibitem{Meng:1996yn}
R.~Meng, F.~I.~Olness and D.~E.~Soper,
Phys.\ Rev.\ D {\bf 54}, 1919 (1996)
[hep-ph/9511311].

\bibitem{yuan}
F.~Landry, R.~Brock, G.~Ladinsky and C.~P.~Yuan,
Phys.\ Rev.\ D {\bf 63} (2001) 013004
[hep-ph/9905391].
C.~Balazs and C.~P.~Yuan,
Phys.\ Rev.\ D {\bf 56}, 5558 (1997)
[hep-ph/9704258].
G.~A.~Ladinsky and C.~P.~Yuan,
Phys.\ Rev.\ D {\bf 50}, 4239 (1994)
[hep-ph/9311341].


\end{thebibliography}
\end{document}